\documentclass[english,12pt]{article}
\makeatletter
\renewcommand\@biblabel[1]{}
\makeatother

\usepackage{array}
\usepackage{braket}
\usepackage{graphicx}
\usepackage{amssymb}
\usepackage{amsmath}
\usepackage{multirow}
\usepackage{prettyref}
\usepackage{babel}
\usepackage{units}
\usepackage[latin1]{inputenc}
\usepackage{amsfonts}
\usepackage{amssymb}
\usepackage{babel}
\usepackage{color}
\usepackage{longtable} 
\usepackage{pdflscape}  
\usepackage{caption} 
\usepackage[numbers]{natbib} 
\usepackage{cite}
\def\@fmsl@sh#1#2#3{\m@th\ooalign{$\hfil#1\mkern#2/\hfil$\crcr$#1#3$}}
 \def\eq#1\en{\begin{equation}#1\end{equation}}
\def\s[#1,#2]{[#1\stackrel{\star}{,}#2]}
\def\sx[#1,#2]{[#1\stackrel{\star_{x}}{,}#2]}

\newcommand{\nc}{\newcommand}
\nc{\beq}{\begin{equation}}
\nc{\eeq}{\end{equation}}
\nc{\beqa}{\begin{eqnarray}}
\nc{\eeqa}{\end{eqnarray}}

\def\bc{\begin{center}}
\def\ec{\end{center}}

\def\gsim{\mathrel{\mathpalette\atversim>}}

\def\bc{\begin{center}}
\def\ec{\end{center}}

\def\gsim{\mathrel{\rlap{\lower4pt\hbox{\hskip1pt$\sim$}}

    \raise1pt\hbox{$>$}}}       %greater than or approx. symbol

\def\gsim{\mathrel{\rlap{\lower4pt\hbox{\hskip1pt$\sim$}}
    \raise1pt\hbox{$>$}}}       %greater than or approx. symbol

\begin{document}
\makeatletter
\def\fmslash{\@ifnextchar[{\fmsl@sh}{\fmsl@sh[0mu]}}
\def\fmsl@sh[#1]#2{%
  \mathchoice
    {\@fmsl@sh\displaystyle{#1}{#2}}%
    {\@fmsl@sh\textstyle{#1}{#2}}%
    {\@fmsl@sh\scriptstyle{#1}{#2}}%
    {\@fmsl@sh\scriptscriptstyle{#1}{#2}}}
\def\@fmsl@sh#1#2#3{\m@th\ooalign{$\hfil#1\mkern#2/\hfil$\crcr$#1#3$}}
\makeatother
%\baselineskip 24pt

%%%%%%%%%%%%%%%%%%%%%%%%%%%%%%%%%%%%%%%%%%%%%%%%%%%%%%%%%%%%%%%%%
%%%
%%%                      TITLE PAGE
%%%
%%%%%%%%%%%%%%%%%%%%%%%%%%%%%%%%%%%%%%%%%%%%%%%%%%%%%%%%%%%%%%%%%
\thispagestyle{empty}
\begin{titlepage}
\boldmath
\begin{center}
  \Large {\bf An analytical perturbative solution to the Merton Garman model using symmetries}
    \end{center}
\unboldmath
\vspace{0.2cm}
\begin{center}
{  {\large Xavier Calmet}\footnote{Corresponding author, e-mail: x.calmet@sussex.ac.uk, phone number: +44  (0)1273 877029, Fax: +44 (0)1273 678097.}$^{a}$ and {\large Nathaniel Wiesendanger Shaw}\footnote{E-mail: nmw24@sussex.ac.uk}$^{a,b}$}
 \end{center}
\begin{center}
$^{a}${\sl School of Mathematical and Physical Sciences, 
University of Sussex, Brighton, BN1 9QH, United Kingdom
} \\
$^{b}${\sl Business School, University of Sussex, Falmer, Brighton BN1 9RH, United Kingdom}
\end{center}
\vspace{5cm}
\begin{abstract}
\noindent
In this paper, we introduce an analytical perturbative solution to the Merton Garman model. It is obtained by doing perturbation theory around the exact analytical solution of a model which possesses a two-dimensional Galilean symmetry. We compare our perturbative solution of the Merton Garman model to Monte Carlo simulations and find that our solutions performs surprisingly well for a wide range of parameters. We also show how to use symmetries to build option pricing models. Our results demonstrate that the concept of symmetry is important in mathematical finance. 
\end{abstract}  
%%%%%%%%%%%%%%%%%%%%%%%%%%%%%%%%%%%%%%%%%%%%%%%%%%%%%%%%%%%%%%%%%
%%%
%%%                     ACKNOWLEDGEMENTS
%%%
%%%%%%%%%%%%%%%%%%%%%%%%%%%%%%%%%%%%%%%%%%%%%%%%%%%%%%%%%%%%%%%%%
\vspace{1cm}

Acknowledgments:
The work of XC is supported in part  by the Science and Technology Facilities Council (grant number  ST/P000819/1). XC is very grateful to MITP for their generous hospitality during the academic year 2017/2018 where most of this work was completed. The authors would like to thank Andreas Kaeck for useful discussions. 

\end{titlepage}

%\pacs{}

%%%%%%%%%%%%%%%%%%%%%%%%%%%%%%%%%%%%%%%%%%%%%%%%%%%%%%%%%%%%%%%%
%%%
%%%                     INTRODUCTION
%%%
%%%%%%%%%%%%%%%%%%%%%%%%%%%%%%%%%%%%%%%%%%%%%%%%%%%%%%%%%%%%%%%%

\newpage

\section{Introduction}
Calculating the price of an option is an important challenge in mathematical finance.  The first attempts in that direction are attributed to  Louis Bachelier who during in his Doctoral thesis, Th\'eorie de la sp\'eculation, published in 1900, considered a mathematical model of Brownian motion and its use for valuing options. This work provided the foundations for the Black Scholes model (Black, Scholes (1973)). However, while the Black Scholes model was a breakthrough in the field, it is widely accepted that it has limitations. In particular, the volatility is treated as a constant which is not very realistic. 

Since the seminal works of Black, Scholes (1973) and Merton (1973), more sophisticated models with a time dependent volatility have been proposed. For example, the affine Heston model (See Heston (1993)), which assumes a time-dependent volatility, with a stochastic process involving the square-root of the stochastic volatility,
and a leverage effect, has been implemented in a large number of empirical studies (Andersen et al. (2002), Bakshi et al. (1997), Bates (2000), Bates (2006), Chernov et al. (2003), Huan and Wu (2004), Pan (2002), Eraker (2004) to name a few). Such models have however limitations and are often modified artificially by combining them with models of jumps in returns and/or in volatility (such as in Jones (2003) and Benzoni (2002)). As a consequence, there is a substantial strand of literature devoted to non-affine volatility models, which note that the popular square-root stochastic volatility model is not very realistic (see e.g. (Eraker et al. (2003) Duan and Yeh (2010), A\"{i}t-Sahalia, Kimmel (2007) Christoffersen et al. (2010), Chourdakis and Dotsis (2011) and Kaeck and Alexander (2012)) to name a few). However, the issue with such models is a general lack of closed form characteristic function, which makes pricing much more challenging. As stated in Chourdakis and Dotsis (2011) when regarding the place of non-affine models and the debate of their tractability against affine models: ``does analytically tractability come at the cost of empirical misspecification?''.  It is a useful endeavor to study non-affine model as we propose in this paper, if an analytical solution for the option pricing formula can be found.

A well-known example of such models is the Merton Garman model (Garman (1976) and Merton (1973)) which is indeed a more realistic model as it allows for a time-dependent volatility and it is not restricted to an affine model for the volatility. However, solving non-affine models is time consuming, as it involves numerical methods. Thus, many practitioners are still using the Black Scholes formula to obtain a fast, albeit not necessarily very reliable, price quote for an option.

The aim of our work is twofold. We will derive an analytical approximative solution to the partial differential equation describing the Merton Garman model which enables one fast calculations of option prices. This requires us to identify a ``symmetric'' version of the model which can easily be solved analytically. One can then reintroduce the symmetry breaking terms of the original Merton Garman differential equation and do perturbation theory around the symmetric solution thereby obtaining an approximative but analytical solution to the original Merton Garman differential equation. We then propose a new approach to model building in option pricing based on the concept of symmetry groups and representation theory. This concept has been extremely successful in modern physics. It is at the origin of all successful models in physics, e.g., in particle physics, cosmology or solid state physics. We note that perturbation theory  has been used in option pricing models (Baaquie (1997), Baaquie (2003), Blazhyevskyi and Yanishevsky (2011), Aguilar (2017), Kleinert and Korbel (2016), Utama and Purgon (2016)) but here we organize perturbation theory around a very specific solution, namely that of the symmetrical model which we will introduce in this paper.

This paper is organized as follows. In section 2, we derive the partial differential equation which describes the Merton Garman model. In section 3, we explain how to reduce the original Merton Garman to a simple, symmetrical, model. We present an exact analytical solution to the symmetrical model.  We then restore the original Merton Garman by reintroducing the symmetry breaking terms and provide an analytical perturbative solution to the Merton Garman model.
In section 4, we compare our solutions to different numerical solutions found in the literature. In section 5, we propose a new approach to model building in mathematical finance. Finally, we conclude in section 6.

\section{The Merton Garman model}

In the Merton Garman model, the price of an option is dependent on the time $t$, the price of the underlying $S$ and the volatility $V$. Both $S$ and $V$ are taken to be time-dependent functions and thus  the Merton Garman model has the potential to provide a more accurate calculation of an option price than e.g. the Black Scholes model.

We start from the stochastic differential equations for the price of the underlying $S$ and for the volatility $V$
\begin{align}
\label{dynamic 1}	dS &= r S dt + \sqrt{V} S dW^{S}, \\
\label{dynamic 2}	dV &= \kappa(\theta - V) dt + \xi V^{\alpha} dW^{V},
\end{align}
which resembles a stochastic, mean reverting, volatility regime.  Here, $\xi$ is the standard deviation of the volatility and $\kappa$ is the speed of mean reversion to the long run variance $\theta$. The interest rate $r$ is assumed to be constant. The model described by Eqs (\ref{dynamic 1}) and (\ref{dynamic 2}) covers many well-known stochastic volatility models, for instance setting $\alpha=1,1/2$ recovers the Hull and White (see Hull and White (1987)) and Heston models, respectively. However, we do not constrain ourselves to either of these worlds. Here, $\alpha$ can take arbitrary values. We will denote the correlation between the two Brownian motions $W^{S}$ and $W^{V}$ by $\rho$.

We shall first consider a call option, but our results can be extended to a put option in a straightforward manner. Our first step is to find the associated partial differential equation  which describes this model. We do so by applying the Feynman-Kac formula, see e.g Hull (1997), which states that for the price of a call option, as defined by the model dynamics in Eqs (\ref{dynamic 1}) and (\ref{dynamic 2}) is given by: 
\begin{align}
	\frac{\partial C}{\partial t} + \sum_{i=1} \mu_{i}(t,x)\frac{\partial C}{\partial x_{i}} + \frac{1}{2}\sum_{i,j=1}\rho_{ij}\sigma_{i}(t,x)\sigma_{j}(t,x)\frac{\partial^{2}C}{\partial x_{i}\partial x_{j}} - rC = 0,
\end{align}
where $C$ is the price of a call.

Using this formula, one obtains
 \begin{eqnarray}
    \label{original MG} \frac{\partial C}{\partial t}+ r S \frac{\partial C}{\partial S} + \frac{1}{2} V S^2 \frac{\partial^2 C}{\partial S^2} +(\lambda + \mu V) \frac{\partial C}{\partial V} +\rho \xi V^{1/2+\alpha} S \frac{\partial^2 C}{\partial S \partial V} +\frac{1}{2}\xi^2 V^{2 \alpha}  \frac{\partial^2 C}{\partial V^2} - rC = 0.
    \end{eqnarray}
where $\lambda = \kappa\theta$, $\mu = -\kappa$. The call price $C=C(S,V,t)$ depends on the time $t$, the price of the underlying $S$ and the time dependent volatility $V=V(t)$. In this model, there are three free parameters $\lambda$, $\mu$ and $\alpha$.  As we explained previously, existing solutions to this partial differential equation are numerical ones which have been obtained using Monte Carlo methods.  Note that the put price $P=P(S,V,t)$ fulfills the same differential equation, but it is obviously subject to a different boundary condition.

\section{Reduction to the symmetrical model and perturbative solution to the Merton Garman model}

By studying the partial differential equation given in Eq. (\ref{original MG}), it quickly becomes clear that the difficulty in finding an analytical solution to Eq. (\ref{original MG}) is due to the lack of symmetry between the different terms of the partial differential equation. It is useful to study the dimensions of the 
different terms and constants in this partial differential equation. The price of the call is obviously given in a specific currency which we shall take to be the USD or $\$$. The remaining dimensions follow from this. We have:
\begin{itemize}
\item $[C]=\$$
\item $[\frac{\partial C}{\partial t}]=\$/\mbox{time}$
\item $[r S \frac{\partial C}{\partial S}]= [r] \$ $ thus $[r]=1/\mbox{time}$
\item $[\frac{1}{2} V S^2 \frac{\partial^2 C}{\partial S^2}]=[V] \$$ thus  $[V]=1/\mbox{time}$
\item $[(\lambda + \mu V) \frac{\partial C}{\partial V}]=([\lambda] + [\mu] 1/\mbox{time})$ thus $[\lambda]=1/\mbox{time}^{2}$ and $[\mu]=\frac{1}{\mbox{time}}$
\item $[\rho \xi V^{1/2+\alpha} S \frac{\partial^2 C}{\partial S \partial V}]= [\rho] [\xi] (1/\mbox{time})^{1/2+\alpha} \$ \mbox{ time}=\$ /\mbox{time}$ thus $[\rho] [\xi]= \mbox{time}^{\alpha-3/2}$
\item $[\xi^2 V^{2 \alpha}  \frac{\partial^2 C}{\partial V^2}]=[\xi]^2 (1/\mbox{time})^{2 \alpha-2} \$ = \$/\mbox{time} $ thus $[\xi]=\mbox{time}^{\alpha-3/2}$ and $\rho$ is dimensionless.
\end{itemize}
It is instructive to see that $S$ and $V$ have different dimensions.  Nevertheless, our goal is to treat $S$ and $V$ as symmetrically as possible to make a global Galilean invariance in 2+1 manifest (see section \ref{Model building in finance, symmetries and group theory}). This can be achieved by adequate variable transformations and by identifying the terms in the differential equation that violate this symmetry.

\subsection{Symmetrical model}
Our aim is to derive a differential equation that is symmetrical in $S$ and $V$. With this aim in mind, let us introduce an averaged volatility $\sigma^{2}$ which is constant. As in the case of the Black Scholes model, different definitions for the averaged volatility are possible, the specific choice will not impact our methodology and results. 

Inspecting the differential equation (\ref{original MG}), it is clear that we need to pick $\alpha=1$  to emphasize the symmetry between $S$ and $V$. We thus consider
\begin{eqnarray}
\frac{\partial C}{\partial t}+ r S \frac{\partial C}{\partial S} + \frac{1}{2} \sigma^{2} S^2 \frac{\partial^2 C}{\partial S^2} +\mu V \frac{\partial C}{\partial V} +\rho \xi_0 V^{3/2} S \frac{\partial^2 C}{\partial S \partial V} +\frac{1}{2}\xi_0^2 V^{2}  \frac{\partial^2 C}{\partial V^2}= r C.
\end{eqnarray}
We need to keep in mind that we will need to reintroduce $\frac{1}{2} V S^2 \frac{\partial^2 C}{\partial S^2}$,  $\lambda \frac{\partial C}{\partial V} $ and the terms corresponding to deviations from 1 for $\alpha$. Note that $\xi_{0}$ is different from $\xi$, in particular they do not have the same dimensions. Finally, we see that there is a mixed derivative term which needs to be eliminated. We thus set $\rho=0$ and we will reintroduce this term as symmetry breaking term.  We thus end up with:
\begin{equation}
\label{symMod} \frac{\partial C}{\partial t} + rS\frac{\partial C}{\partial S} + \frac{1}{2}{\sigma^{2}}S^{2}\frac{\partial^2 C}{\partial S^2} + \mu V \frac{\partial C}{\partial V} +\frac{1}{2} \xi_0^2 V^2 \frac{\partial^2 C}{\partial V^2} =r C.
\end{equation}
This partial differential equation can be massaged with standard substitutions into a 2+1 dimensional heat equation (see Appendix \ref{Heat equation}) in which case the symmetry in $S$ and $V$ becomes manifest. In order to do so, we introduce
\begin{align}
\label{xy substitutions}	
x &= \log(S/K), \\
y &= \log(V/V_{0}),
\end{align}
and
\begin{equation}
\label{phi substitution} C(x,y,\tau) = K \phi(x,y,\tau)\psi_{0}(x,y,\tau),
\end{equation}
where $K$ is the strike price and $V_{0}$ is some constant with units of $1/\text{sec}$. The function $\phi(x,y,\tau)$ and the rescaled time $\tau$ are defined in Appendix \ref{Heat equation}. Standard manipulations described in Appendix \ref{Heat equation} lead to
\begin{equation}
\label{Heat} \frac{\partial \psi_{0}}{\partial \tau} = \frac{\partial^2 \psi_{0}}{\partial x^{2}} + \frac{\partial^2 \psi_{0}}{\partial y^2}.
\end{equation}
which is manifestly symmetrical in $x$ and $y$. We will thus refer to the model described by the differential equation (\ref{symMod}) as the symmetrical model.  Another reason for massaging the symmetrical model into a heat equation is that this equation is easy to solve analytically. 

We impose the standard boundary condition for the call price:
\begin{equation}
C(S,V,T) = \Bigg{(}  S(T)- K  \Bigg{)}^{+}.
\end{equation}
For a put option we have
\begin{equation}
P(S,V,T) = \Bigg{(} K- S(T)  \Bigg{)}^{+}.
\end{equation}

\subsection{Solution of the symmetrical model}
Details of the derivation of the analytical solution of the symmetrical model, i.e., of the 2+1 dimensional heat equations, are given in Appendix \ref{Solution of the symmetrical model}. We find
\begin{equation}
C_{0}(S,V,t) = S \mathcal{N}(d_1) -K e^{-r(T-t)}\mathcal{N}(d_2),
\end{equation}
where
\begin{align}
\mathcal{N}(d)=\frac{1}{\sqrt{2 \pi}} \int_{-\infty}^d \exp\left(\frac{-z^2}{2}\right) dz,
\end{align}
and
\begin{align}
d_{1} &= \frac{x}{\sqrt{2\tau}} + \frac{\sqrt{2\tau}}{2}(R_{1} + 1)=\frac{\log(S/K)+(r+\sigma^2/2)(T-t)}{\sigma \sqrt{T-t}}, \\
d_{2} &= \frac{x}{\sqrt{2\tau}} + \frac{\sqrt{2\tau}}{2}(R_{1} - 1) = d_1-\sigma\sqrt{T-t}.
\end{align}
 Remarkably, because of the boundary condition that only depends on $S$, it is identical to the Black Scholes solution.  
 
 For a put option, the very same procedure leads to
 \begin{equation}
P_{0}(S,V,t) = C_{0}(S,V,t)-S+K e^{-r(T-t)}.
\end{equation}
 In the next subsection, we shall restore the symmetry breaking terms and discuss the full Merton Garman model.

\subsection{Symmetry Breaking terms and solution to the Merton Garman model}

We are now in a position to solve the full Merton Garman model using perturbation theory around the symmetrical solution $C_{0}(S,V,t)$. We organize perturbation theory as an expansion in terms the coefficients of the symmetry breaking terms.  We first need to restore the full model by re-introducing the symmetry breaking terms
\begin{multline}
\label{fullM}	\frac{\partial C}{\partial t} + rS\frac{\partial C}{\partial S} + \frac{1}{2}\sigma^{2}S^{2}\frac{\partial^{2}C}{\partial S^{2}} + \frac{c_{1}S^{2}}{2}\Bigg{(} V - \sigma^{2} \Bigg{)} \frac{\partial^{2}C}{\partial S^{2}} + \mu V\frac{\partial C}{\partial V} + c_{2}\lambda\frac{\partial C}{\partial V} + \frac{1}{2} \xi_{0}^{2}V^{2}\frac{\partial^{2}C}{\partial V^{2}}  
	\\ + c_{3} \frac{1}{2} \Bigg{(} \xi^{2}V^{2\alpha} - \xi_{0}^{2}V^{2}  \Bigg{)}\frac{\partial^{2} C}{\partial V^{2}} + c_{4}\rho\xi V^{\alpha + 1/2}S\frac{\partial^{2} C}{\partial S\partial V} - r C = 0.
\end{multline}
Note that we have introduced dimensionless coefficients $c_{i}$ which denote the strength of the symmetry breaking terms. In the limit $c_{i}=1$ one recovers the original Merton Garman model. These coefficients are simply introduced as a bookkeeping trick to keep track of which terms correspond to a deviation of the 2+1 Galilean invariant theory. In the end of the day, we set $c_{i}=1$.  We now do perturbation theory around the symmetrical solution $C_{0}(S,V,t)$ and obtain
 \begin{eqnarray} 
C_1(S,V,t)&=&-K  \frac{\left(\frac{S}{K}\right)^{\frac{1}{2}-\frac{r}{\sigma ^2}} e^{\left(\frac{4 \log
   ^2\left(\frac{S}{K}\right)+\left(2 r+\sigma ^2\right)^2 (t-T)^2}{8 \sigma ^2
   (t-T)}\right)} }{4 \sqrt{2 \pi } \left(\frac{2 \gamma }{\sigma \xi_{0}}+1\right) \sqrt{\sigma ^2 (T-t)}} \\ \nonumber 
   &&  \times \left(\frac{1}{2} \sigma ^4 \left(\frac{2 \gamma }{\sigma \xi_{0}}+1\right) (t-T)+V \left(e^{\frac{1}{2} \sigma ^2 \left(\frac{2 \gamma }{\sigma \xi_{0}}+1\right) (T-t)} - 1 \right)\right).
 \end{eqnarray}
where we have set $c_1=1$. We expect that our approximation should work well when $\lambda$ and $\rho$ are small, when $\alpha$ is close to one and when the variation of $V$ around is average value $\sigma^2$ is not too large. In the limit when $V$ is large, $\sigma^2$ is large as well and we expect that, as in the Black Scholes case, the price of the call becomes the price of the underlying $S$. Details of the derivation can be found in Appendix \ref{Symmetry Breaking terms and solution to the Merton Garman model}. 

It may appear surprising that the leading order correction does not depend on the symmetry breaking terms parametrized by $c_2$, $c_3$ and $c_4$. It can easily be shown (see Appendix \ref{Symmetry Breaking terms and solution to the Merton Garman model}) that the boundary condition  (\ref{transformed BC}) insures that only the contribution from the $c_1$ term survives. The boundary condition implies that the contributions of $c_2$, $c_3$ and $c_4$ vanish to leading order in the perturbation theory. These symmetry breaking terms will, however, contribute to higher order corrections. Higher precision, if required, can be obtained by going to higher order in perturbation theory. Option prices can be calculated extremely rapidly using this formalism. Note that, in principle, if we resummed perturbation theory to all order in $c_i$, the dependence on $\sigma$ and $\xi_0$ would  vanish. It is also worth noticing that our results are independent on $V_0$ which is only introduced to match the dimension of $V$. 

It is straightforward to show that we obtain the same result for a put option
   \begin{eqnarray} 
P_1(S,V,t)&=&-K  \frac{\left(\frac{S}{K}\right)^{\frac{1}{2}-\frac{r}{\sigma ^2}} e^{\left(\frac{4 \log
   ^2\left(\frac{S}{K}\right)+\left(2 r+\sigma ^2\right)^2 (t-T)^2}{8 \sigma ^2
   (t-T)}\right)} }{4 \sqrt{2 \pi } \left(\frac{2 \gamma }{\sigma \xi_{0}}+1\right) \sqrt{\sigma ^2 (T-t)}} \\ \nonumber 
   &&  \times \left(\frac{1}{2} \sigma ^4 \left(\frac{2 \gamma }{\sigma \xi_{0}}+1\right) (t-T)+V \left(e^{\frac{1}{2} \sigma ^2 \left(\frac{2 \gamma }{\sigma \xi_{0}}+1\right) (T-t)} - 1 \right)\right).
 \end{eqnarray}

The prices obtained to leading order in perturbation theory for a call and put option are thus given by
\begin{equation}
C(S,V,t) = C_{0}(S,V,t)+C_{1}(S,V,t),
\end{equation}
and
\begin{equation}
P(S,V,t) = P_{0}(S,V,t)+P_{1}(S,V,t).
\end{equation}
In the next section we shall compare our results to exact solutions obtained numerically.

\section{Comparison with numerical simulations}

In this section we investigate how the approximative solution compares to a Monte Carlo simulation of the full Merton Garman model. 
It is well-known that the Merton Garman model is not solvable analytically, however it can be solved using numerical methods. The first step is a comparison of static cross sections of options. Then we compare using a simulated time series calibration exercise.
\subsection{Static Cross-Section Comparison}

The first step in evaluating the performance of the leading order perturbative solution is to compare to a multitude of simulated data of the Merton Garman model to ensure that the approximation is sufficient to fit a range of different options at one time. We start by describing the data simulation of the Merton Garman model, then move onto the calibration procedure for the approximative solution and discuss the results. 

We choose a standard Monte Carlo framework, using stratified sampling and antithetic variables, simulating seven million paths with a time step of one-tenth of a day for option maturities. We choose a spot underlying price of $S=\$100$ and a strike range of $K\in [90,110]$ to give a moneyness range of $K/S\in[0.9,1.1]$ to simulate call options throughout the spectrum of moneyness \footnote{In this exercise we simulate call options only, as put prices can be calculated from the put-call parity.}. We simulate the Merton Garman model with the structural parameter vector: $\Theta^{MG}=\{1.5,0.08,1.5,-0.5,1\}$ \footnote{We also simulate for $\alpha\in [0.75,1.5]$ and $\kappa \in [1.5,5], \rho \in [-0.5,-0.9], \theta\in [0.08,0.15]$. However, the results of the calibration exercise are represented by the choice of parameters made above.} and initial volatility $V(t=0) \in [10,35] \%$.

The leading order perturbative solution is independent on the the symmetry breaking terms, characterized by $c_{2}$, $c_{3}$ and $c_{4}$. The solution is thus independent of the parameters: $\rho,\theta$. However, as a by-product of the perturbation theory we introduced the following parameters: $\xi_{0},\sigma$. From inspection we fix $\xi_0$ using  $\xi_{0} = \xi\sigma^{2(\alpha-1)}$ which guarantees that it has the right dimensions. While $\sigma$ remains to be determined by calibration. Yielding the parameter vector: $\Theta^{pert.}=\{\kappa,\xi,\alpha,\sigma\}$.

\begin{figure}[!htb]
	\caption{Comparative fit of four different initial volatilities. Odd numbered panels (1,3,5,7) display the natural logarithm difference of the prices, while even numbered panels (2,4,6,8) display the implied volatility curves. For the implied volatility panels the solid black line represents the Monte Carlo implied volatility and the dashed line is that of the leading order order perturbative solution. We pick an option maturity of 30 days.}
	\centering
	\includegraphics[scale=0.70]{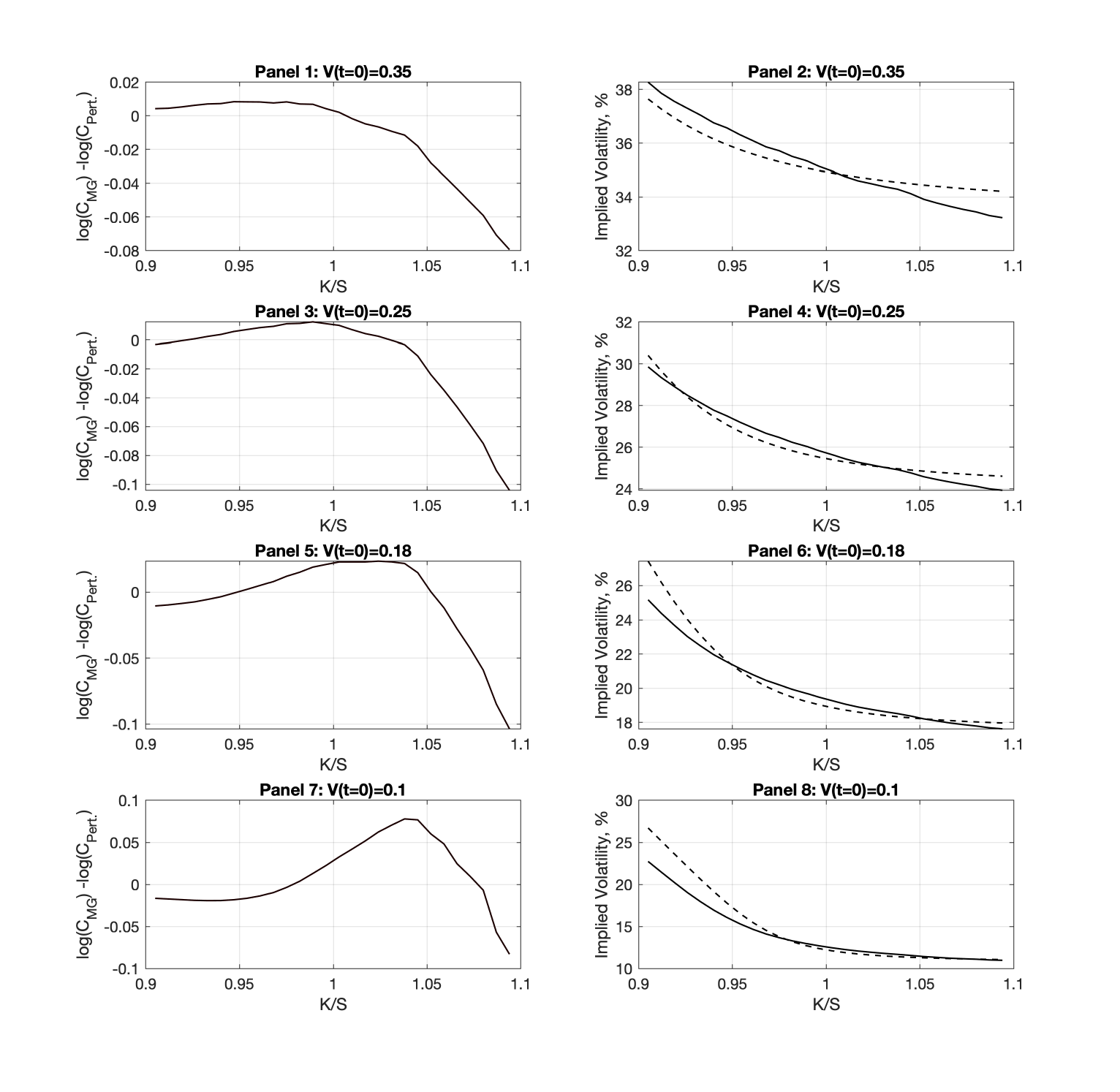}
	\label{fig: static}
\end{figure}

The parameter $\sigma$ is determined by calibration. When fitting $\sigma$ to these simulations, there is a risk of overfitting expensive out of the money (OTM) options. For that reason, it is best to consider the implied volatility objective function (this is noted in Christoffersen et al. (2014)) which is given by: 
\begin{equation}
\label{IV objective function} IVRMSE = \sqrt{\frac{1}{N}\sum^{N}_{i=1}(IV_{MC_{i}} - IV_{pert._{i}})^{2}},
\end{equation}
where $IV_{MC_{i}}$ stands for the implied volatility of the $i^{\text{th}}$-option simulated using the Monte Carlo and $IV_{pert._{i}}$ is the implied volatility of the $i^{\text{th}}$-option calculated using the leading order perturbative solution. The calibration exercise is extremely fast as our formula for option prices is an approximative analytical solution. Figure (\ref{fig: static}) and Table \ref{tab: static} demonstrates the results of the static calibration exercise for a 30 day maturity horizon \footnote{While we simulated time horizons between 5-100 days, this is representative of our results and we drop the other results for brevity.}. It should be noted that in Figure (\ref{fig: static}) the price Panels are the difference of the log, this is needed to observe any difference in prices as the two methods produce prices which are very similar. However, from these Panels it is clear that smaller moneyness, i.e $K/S<1$ see extremely small errors where $\log(C_{MC}/C_{pert.}) \sim 0$. Also apparent is that at some moneyness level the leading order perturbative solution will over price options, the level of moneyness at which this occurs is inversely proportional to volatility. Lastly, the range of under to over pricing is also inversely proportional to volatility. However, as the prices from both methods are very similar it is more informative to look at the implied volatility cross-section in the even Panels of Figure (\ref{fig: static}) along with the IVRMSE column of Table \ref{tab: static}. These show throughout the range of volatility regimes the leading order perturbative solution is able to approximate successfully a low IVRMSE, with a maximum occurring from the low volatility regime (Panel 8) of $1.57\%$.  

\begin{figure}[!htb]
	\caption{Test of the validity of the perturbation theory. These panels depict the ratios $C_1/(C_0+C_1)$ in  $\%$ where $C_0$ is the contribution to the price of the symmetrical solution and $C_1$ the leading order correction in perturbation theory. Clearly $C_1 \ll C_0$ even when the volatility is large. This demonstrates the validity of the perturbative expansion. The four cases correspond respectively to panels (1,3,5,7) of Figure 1.}
	\centering
	\includegraphics[scale=0.60]{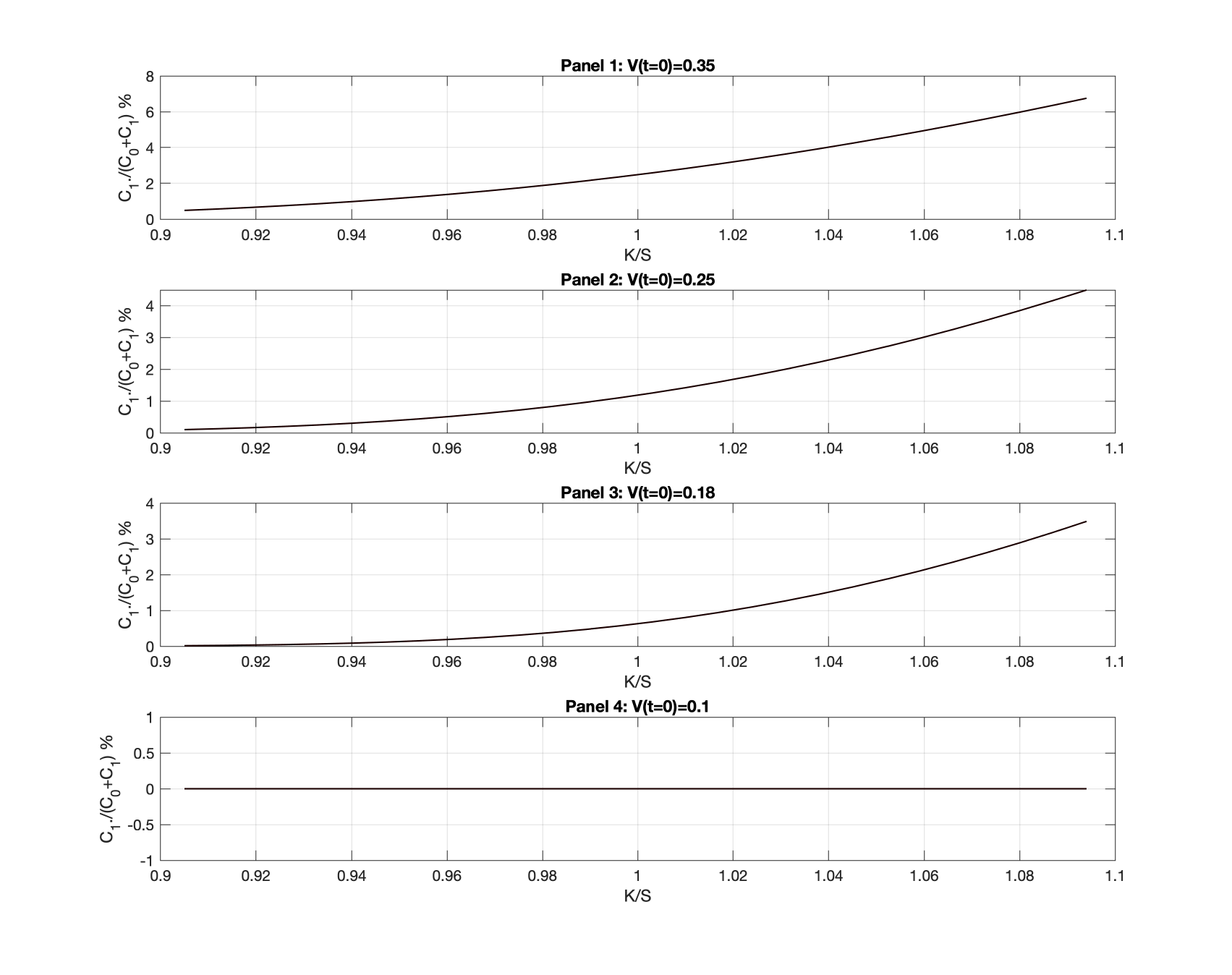}
	\label{fig: perttheory}
\end{figure}

Another way to test the consistency of the perturbative expansion is to consider the ratio $C_1/(C_0+C_1)$  as a function of moneyness, $K/S$ , where $C_0$ is the contribution to the price of the symmetrical solution and $C_1$ is the leading order correction in perturbation theory. Figure (\ref{fig: perttheory}) shows these ratios for several cases. Clearly $C_1 \ll C_0$ even when the volatility is large. This demonstrates nicely the validity of the perturbative expansion even in the case where the volatility is large.  While this exercise confirms that for fixed scenarios the perturbative solution is a very good approximation to the actual Merton Garman solution, it is essentially a multiple curve fitting exercise, a more thorough analysis is needed to be able to gauge the reliability of the perturbative approximation. This is what we shall focus on next.

\subsection{Simulated Time Series Calibration}
The second step in evaluating the performance of the leading order perturbative solution is to estimate it against a time series simulation of the Merton Garman model. The time series uses 100 different Monte Carlo paths to simulate the asset price and variance paths including 6 unique maturities within $[7,180]$ days, for details see Appendix \ref{appendix: time series}.

The benefits of the stress test is two fold: firstly it is particularly pertinent to run a number of different simulations for different parameter values, specifically investigating the effect different $\theta,\rho$ has on performance, as the other parameters in the $\Theta^{\text{pert.}}$ vector will have to attempt to absorb the information contained in the absent parameters. Secondly, it also provides a first estimate into the applicability of the perturbative solution to different types of options markets. We simulate four different data sets with varying parameter vectors, described below.

\begin{itemize}
 \item \textit{data set 1}: $\Theta^{MG}= \{1.1768, 0.0823, 0.3000, -0.5459,1.0000\}$, with a negative correlation which is a reasonable choice for modeling equity options, such as the S$\&$P 500. 

\item \textit{data set 2}: $\Theta^{MG}= \{1.1768, 0.0823, 0.3000, 0.0000,1.0000\}$, with a correlation of zero. 

\item \textit{data set 3}: $\Theta^{MG}= \{1.1768, 0.0823, 0.3000, +0.5459,1.0000\}$, with a positive correlation coefficient this is used to gain inference about modeling VIX options, see Park (2015).

\item \textit{data set 4}: $\Theta^{MG}= \{1.1768, 0.1250, 0.3000, -0.5459,1.0000\}$, with a high(er) central tendency we investigate an equity style option market with a central tendency which is significantly higher than the initial variance value of: $0.08$ and thus tests how the leading order perturbative solution handles significant change in the variance path. 
\end{itemize}

For the following simulated calibration exercises, it is imperative to note the difference in IVRMSE and parameters between the data sets as this will highlight the following: firstly, parameter regions where the leading order perturbative solution might breakdown. Secondly, potential difficulties in estimating certain parameters of the model. Thirdly, where information contained in $\theta,\rho$ might be absorbed. These results can be found in Tables \ref{tab: parameter results}-\ref{tab: error results}. Table \ref{tab: parameter results} contains the results of the parameter vector estimates for data sets 1-4 along with summary statistics comparing to the Merton Garman  parameter vector. Table \ref{tab: error results} contains results of the IVRMSE and standard deviations for each data set. The results of the data sets are described below:

\begin{itemize}
 \item  \textit{Data set 1}: from Table \ref{tab: parameter results} Panel 1 parameters $\kappa,\xi$ appear to be challenging to estimate, being significantly larger and with quite high standard deviations, while $\alpha$ appears to be stable. While $\Theta^{\text{Pert.}}$ does not contain $\theta$ a significant amount of this missing information is absorbed by $\sigma$ and $\xi$. Table \ref{tab: error results} demonstrates the leading perturbative solution does very well in approximating the Merton Garman model with an IVRMSE of $1.2977\%$ and standard deviation of $0.3474\%$.
\item \textit{Data set 2}: from Table \ref{tab: parameter results} Panel 2 it starts to become clear some of the information contained in the correlation coefficient is absorbed by both $\xi,\alpha$, particularly the latter. With the value of $\alpha$ reducing significantly while the standard deviation approximately doubles (relative to data set 1). Although it does appear that this regime is slightly easier to estimate $\kappa,\xi$. Table \ref{tab: error results} reports a significant decrease (relative to data set 1) in IVRMSE with a moderate decrease in standard deviation.
\item \textit{Data set 3}: Table \ref{tab: parameter results} Panel 3 demonstrates the absence of the information contained in the correlation coefficient has an effect on the ability to estimate $\kappa,\xi,\alpha$ in a similar manor to that of Panel 1, observing very similar biases and standard deviations. Perhaps suggesting that it is more the non-zero nature of the correlation coefficient which the leading perturbative solution struggles with. Furthermore, Table \ref{tab: error results} demonstrates very similar errors to data set 1.
\item \textit{Data set 4}: from Table \ref{tab: parameter results} Panel 4 the increase in central tendency also clearly has an impact in $\xi,\alpha$ the two variance related parameters of the leading perturbative solution. This is due to the increased difference in magnitude between the central tendency and initial variance. It suggests that both parameters absorb missing information contained in $\theta$. This difference manifests itself in Table \ref{tab: error results} with a resulting IVRMSE of $1.7199\%$, the largest across all data sets. 
\end{itemize}

In summary, on the estimation side $\kappa$ could certainly be a challenge to estimate, although this is not unique to our approach. For substantial difference between variance and central tendency it appears that $\alpha,\xi$ could also be a challenge, other than this estimating $\alpha$ is generally inconsiderable. Regarding the matter of information absorption we note that it is clear that $\sigma$ absorbs significant amount of the information contained in $\theta$, with contribution from $\alpha$ for large disparity between initial variance and central tendency. The parameters $\xi,\alpha$ also seem to share the majority of the information contained in $\rho$. Table \ref{tab: error results} indicates that across data sets the leading perturbative solution does well in approximating the Merton Garman model with fairly consistent errors, given the standard deviations, with an approximate error range of $1.2-1.7\%$.

\section{Model building in finance, symmetries and group theory}
\label{Model building in finance, symmetries and group theory}

In this section, we shall first discuss Galilean invariance, see e.g. (Bose (1995) and L\'evy-Leblond (1967)), in the context of mathematical finance before explaining how new option pricing models can be constructed using the concept of symmetries. 
We shall assume that the dimensionless option price $\psi(x,y,t)$, from which the usual price of the option is derived,  is the fundamental quantity.  If we posit that $\psi(x,y,t)$ is a measurable quantity, it should not depend on the coordinate system ${\cal P}$ that is being used to measure it. We could use ${\cal P}$ parametrized by $(x,y,t)$ or ${\cal P}^\prime$ parametrized by $(x^\prime,y^\prime,t^\prime)$ and obtain the same dimensionless price, assuming that the two coordinate systems are related by a transformation which we shall take to be a Galilean transformation, knowing that $\psi(x,y,t)$ is a solution to the 2+1 heat equation.
A Galilean transformation can be decomposed as the composition of a rotation, a translation and a uniform motion in the space $(x,y,t)$ where $\vec x=(x,y)$ represents a point in two-dimensional space, and $t$ a point in one-dimensional time. A general point in this space can be ordered by a pair $(\vec x, t)$. Let us first consider the case where the two coordinate systems are moving away from each other in a uniform motion fixed by a two-dimensional constant vector $\vec w$.
A uniform motion  with vector $\vec w$, is given by
\begin{eqnarray}
(\vec x, t)\mapsto (\vec x+ t \vec w, t).
\end{eqnarray}
 A translation is given by
\begin{eqnarray}
(\vec x, t)\mapsto (\vec x+\vec a, t+ s),
\end{eqnarray}
where $\vec a$ is two-dimensional vector and $s$ a real number. Finally a rotation is given by
\begin{eqnarray}
(\vec x, t)\mapsto (G \vec x, t),
\end{eqnarray}
where $G$ is $2\times 2$ orthogonal transformation. It is easy to see that the partial differential equation obeyed by $\psi(x,y,\tau)$
\begin{equation}
	\left( \frac{\partial }{\partial \tau} -  \frac{\partial^2 }{\partial x^2}- \frac{\partial^2 }{\partial y^2}\right)\psi(x,y,\tau)=0,
\end{equation}
is covariant under two-dimensional Galilean transformations Gal(2).

Our Monte Carlo investigation of the Merton Garman model demonstrates that the symmetry Gal(2) is a good symmetry of the model as the leading perturbation theory around the symmetrical solution works very well. This demonstrates the importance of the symmetry between $S$ and $V$ which is manifest when expressed in the appropriate variables. To a very good approximation and for a range of parameters relevant to financial applications, the Merton Garman model possesses a hidden Gal(2) symmetry that is only softly broken. This is very clearly illustrated by the results obtained in Figure (\ref{fig: perttheory}) which demonstrates that the larger the volatility fluctuations are, the more the symmetry breaking terms become important. In panel 4 where the volatility is very close to being constant, the leading order correction is essentially vanishing and the symmetrical solution is very close to that obtained with the original Merton Garman model. This is suggestive of an alternative way for building option pricing models. Instead of starting from stochastic processes, we could simply have derived the symmetrical model by positing that the option price should depend on the price of the underlying, a time dependent volatility and time. By requiring that the dimensionless option price follows a differential equation that is Gal(2) covariant. We would immediately have obtained the 2+1 dimensional heat equation. The very small deviations from the Gal(2) covariance, can be accounted for by symmetry breaking terms as explained above. This led us to an approximative perturbative analytical solution of the original Merton Garman differential equation. 

There is another interesting consequence of having a symmetry group as a fundamental building block. Namely, one can classify how the different objects in the model will transform under this symmetry. This is a very well developed field of mathematics called representation theory. In the case above, $\psi(x,y,t)$ is a scalar under Galilean transformations. A scalar representation of a symmetry group means that it is invariant under the group transformations. The differential equation is covariant under such transformations. Besides the scalar transformations, there are vector representations such as e.g. $\partial \psi(x,y,t)/\partial x$, $\partial \psi(x,y,t)/\partial y$ or $\partial \psi(x,y,t)/\partial t$ which are nothing but the Greeks. They appear from a very different perspective than is usually the case in finance.

These ideas open up new directions in model building for option pricing. One could for example consider a 3+1 dimensional model $(x,y,z,t)$ where the $z$-direction could describe a time dependent interest rate or additionally this could be used to model a two-factor volatility process with stochastic central tendency, see e.g. Bardgett et al. (2018). The differential equation for the price would be of the type
\begin{equation}
 \frac{\partial \psi(x,y,t)}{\partial t} = \frac{\partial^2 \psi(x,y,z,t)}{\partial x^{2}} + \frac{\partial^2 \psi(x,y,z,t)}{\partial y^2}+ \frac{\partial^2 \psi(x,y,z,t)}{\partial z^2},
\end{equation}
for which it is easy to find solutions:
\begin{equation}
G(x,y,\tau;x^\prime,y^\prime, z^\prime,t) =\Theta(\tau-t) \frac{1}{(4 \pi (\tau-t))^{3/2}}
 \exp\left(- \frac{(x-x^\prime)^2+(y-y^\prime)^2+(z-z^\prime)^2}{4(\tau-t)}\right).
\end{equation}
One could also consider ``relativistic'' extensions of the Merton Garman model treating time on the same footing as the underlying price and the volatility:
\begin{equation}
 \frac{\partial^2 \psi(x,y,t)}{\partial t^2} = \frac{\partial^2 \psi(x,y,t)}{\partial x^{2}} + \frac{\partial^2 \psi(x,y,t)}{\partial y^2}.
\end{equation}
It is well known that this model is covariant under Lorentz transformations. Clearly identifying the right symmetry group for a given financial system is of paramount importance. Making use of symmetries to model physical system has been extremely successful in all fields of physics. Applying these ideas to option pricing models opens up new perspectives for model building in finance using the concept of symmetry groups and representation theory.

\section{Conclusions}
In this paper we have introduced a perturbative method to obtain analytical approximative solutions to models such as the Merton Garman model. The key idea consists in treating the price of the underlying and the volatility in a symmetrical way. This leads to a model which has an exact Galilean invariance in two-dimensions as it is described by the two-dimensional heat equation which has an analytical solution. By folding this solution with the boundary condition leading to the correct price at maturity for a call option, we obtained an analytical  symmetrical solution for this model which corresponds to the Black Scholes solution, despite being derived from a very different perspective and in a framework with a time dependent volatility.  

The Merton Garman model is recovered by introducing symmetry breaking terms and we have calculated the leading order correction to the symmetrical solution. We have shown that our perturbative solution works very well by comparing it to a Monte Carlo simulation of the Merton Garman model for a range of parameters which are relevant from a financial point of view. The moneyness curves of the two prices are so overlapping that we had to plot implied volatility curves to be able to discuss in a quantitative manner the differences between the two solutions. 

We argued that the fact that the symmetrical model works so well is a sign that the Merton Garman model has a hidden two-dimensional Galilean symmetry which is softly broken for the relevant parameter ranges.  We have explained that the concept of symmetry, groups and representation theory could be extremely useful in building pricing models in financial mathematics. This clearly needs to be explored further. From this point of view, our work is opening up a new perspective on model building in mathematical finance. 

\vspace{1cm}
{\it Data Availability Statement:} Data sharing is not applicable to this article as no new data were created or analyzed in this study.

\appendix

\section{Heat equation}
\label{Heat equation}

In this Appendix, we show how to massage the partial differential equation corresponding to the symmetrical model:
\begin{equation}
\label{symMod} \frac{\partial C}{\partial t} + rS\frac{\partial C}{\partial S} + \frac{1}{2}{\sigma^{2}}S^{2}\frac{\partial^2 C}{\partial S^2} + \mu V \frac{\partial C}{\partial V} +\frac{1}{2} \xi_0^2 V^2 \frac{\partial^2 C}{\partial V^2} =r C,
\end{equation}
into the heat equation in 2+1 dimensions. To do so, we now make the standard substitutions for the underlying and variance, transforming them to dimensionless variables:
\begin{align}
\label{xy substitutions}	
x &= \log(S/K), \\
y &= \log(V/V_{0}),
\end{align}
where $K$ is the strike price and $V_{0}$ is some constant with units of $1/\text{sec}$. Re-casting Eq. (\ref{symMod}) in terms of $x$ and $y$, we obtain
\begin{equation}
\frac{\partial C}{\partial t} + \omega \frac{\partial C}{\partial x} + \gamma \frac{\partial C}{\partial y} + \frac{1}{2}{\sigma^{2}}\frac{\partial^2 C}{\partial x^2} + \xi_0^2 \frac{\partial^2 C}{\partial y^2} = r C,
\end{equation}
where, $\omega = r - \frac{1}{2}{\sigma^{2}}$ and $\gamma= \mu - \xi_0^2/2$.

In order to remove the constant term, $\frac{1}{2}{\sigma^{2}}$ in front of the second derivative with respect to $x$, we make the time transformation: $\tau = \frac{{\sigma^{2}}}{2} (T-t)$, yielding: 
\begin{equation}
-\frac{\partial C}{\partial \tau} + (R_{1} - 1)\frac{\partial C}{\partial x} + \frac{2\gamma}{{\sigma^{2}}}\frac{\partial C}{\partial y} + \frac{\partial^2 C}{\partial x^2} + \eta\frac{\partial^2 C}{\partial y^2} - R_{1}C = 0,
\end{equation}
where $R_{1} = \frac{2r}{{\sigma^{2}}}$ and $\eta = \frac{2\xi_0^2}{{\sigma^{2}}}$. We then proceed with the further substitution: $y = \frac{1}{\sqrt{\eta}}y$, transforming the coefficient of the second derivative  with respect to $y$ to unity:
\begin{equation}
\label{pre-exponential} -\frac{\partial C}{\partial \tau} + (R_{1} - 1)\frac{\partial C}{\partial x} + \sqrt{2}\frac{\gamma}{\sigma\xi_0}\frac{\partial C}{\partial y} + \frac{\partial^2 C}{\partial x^2} + \frac{\partial^2 C}{\partial y^2} - R_{1}C = 0~.
\end{equation}

Next, we make the price transformation:
\begin{equation}
\label{phi substitution} C(x,y,\tau) = K \phi(x,y,\tau)\psi_{0}(x,y,\tau),
\end{equation}
with: 
\begin{equation}
\phi(x,y,\tau) = e^{a x + b y + c \tau}.
\end{equation}
The constants $a, b$ and $c$ are chosen by inspection, after substitution into  Eq. (\ref{pre-exponential}) we see that the choice: 
%\begin{equation}
\begin{align} \label{a}
a &= -\frac{1}{2}(R_{1} - 1), \\
\label{b}
b &= -\frac{1}{2}(R_{2} - 1), \\
c &= -\frac{1}{4}\Bigg{(}  (R_{2} - 1)^{2} + (R_{1} + 1)^{2} \Bigg{)},
\end{align}
where: $R_{2} = 1 + \sqrt{2}\gamma/\sigma\xi_0$, leads to the heat equation in 2+1 dimensions:
\begin{equation}
\label{Heat} \frac{\partial \psi_{0}}{\partial \tau} = \frac{\partial^2 \psi_{0}}{\partial x^{2}} + \frac{\partial^2 \psi_{0}}{\partial y^2}.
\end{equation}
There are well known techniques to solve this partial differential equation analytically. It is also well known that the heat equation has a symmetry based on the Galilean group in 2+1 dimensions. This symmetry is now manifest. In particular, the variable $x$ and $y$ are interchangeable as advertised previously. Before we can solve the heat equation, we need to specify an appropriate boundary equation.

Guided by our intuition formed by the Black Scholes model, we propose a boundary condition of the form for a call option:
\begin{equation}
\label{transformed BC} \psi_{0}(x,y,0) =    e^{\frac{1}{2}(R_{2} - 1)y}  \Bigg{(}    e^{(R_{1} + 1)x/2 } -  e^{(R_{1} - 1)x/2}  \Bigg{)}^{+}.
\end{equation}
To verify that this boundary condition makes sense from a option pricing point of view, we work out the implied boundary conditions for the call price in the original variables:
\begin{eqnarray}
C(x,y,0) &=&K e^{a x + b y}\psi_{0}(x,y,0), 
\end{eqnarray}
and thus
\begin{eqnarray}
C(x,y,0) &=& K \Bigg{(}  e^{x} - 1  \Bigg{)}^{+}.
\end{eqnarray}
Substitution back to original variable, using: $S = Ke^{x}$, we find
\begin{equation}
C(S,V,T) = \Bigg{(}  S(T)- K  \Bigg{)}^{+}.
\end{equation}
This is the standard payoff of a call option. We have thus found an appropriate boundary condition for the heat equation and can thus proceed to solving this partial differential equation.
For a put option we have
\begin{equation}
\label{transformed BC2} \psi_{0}(x,y,0) =   -e^{\frac{1}{2}(R_{2} - 1)y}  \Bigg{(}    e^{(R_{1} + 1)x/2 } -  e^{(R_{1} - 1)x/2}  \Bigg{)}^{+},
\end{equation}
which leads to 
\begin{equation}
P(S,V,T) = \Bigg{(} K- S(T)  \Bigg{)}^{+}.
\end{equation}

\section{Solution of the symmetrical model}
\label{Solution of the symmetrical model}

In this Appendix, we show how to solve the heat equation in 2+1 dimensions:
\begin{equation}
 \frac{\partial \psi_{0}}{\partial \tau} = \frac{\partial^2 \psi_{0}}{\partial x^{2}} + \frac{\partial^2 \psi_{0}}{\partial y^2}.
\end{equation}
So far the function $\psi_{0}(x,y,\tau)$ is only defined for $\tau > 0$, however by introducing the Heaviside function $\Theta(\tau)$ we may extend the definition domain to  the range $\tau < 0$
\begin{equation}
\bar{\psi}_{0}(x,y,\tau) = \Theta(\tau)\psi_{0}(x,y,\tau),
\end{equation}
Thus we get an inhomogeneous differential equation:
\begin{equation}
\Bigg{(}  \frac{\partial}{\partial \tau} - \frac{\partial^2}{\partial x^2} - \frac{\partial^2}{\partial y^{2}}  \Bigg{)}\bar{\psi}_{0}(x,y,\tau) = \bar{\psi}_{0}(x,y,0) \delta(\tau).
\end{equation}
This equation is solved by
\begin{equation}
\bar{\psi}_{0}(x,y,\tau) = \int \bar{\psi}_{0}(x^\prime,y^\prime,0) G(x,y,\tau| x^\prime,y^\prime,0) dx^\prime dy^\prime.
\end{equation}
This is the partial differential equation for the Gaussian propagator of the heat equation in 2+1 dimensions:
\begin{equation}
G(x,y,\tau | X,Y,0) = \frac{1}{4 \pi\tau} e^{  -\frac{(x-X)^2}{4\tau} -\frac{(y-Y)^2}{4\tau}}  \Theta(\tau).
\end{equation}

Combining the above two results, the solution can be written in the form
\begin{align}
\bar{\psi}_{0}(x,y,\tau) &= \frac{1}{4 \pi\tau} \int \bar{\psi}_{0}(X,Y,0) e^{  -\frac{(x-X)^2}{4\tau} -\frac{(y-Y)^2}{4\tau}} dXdY, 
\end{align}
\begin{multline}
\label{psi zero}	\bar{\psi}_{0}(x,y,\tau) = \frac{1}{4\pi\tau} \int  e^{(R_{2} - 1)Y/2}  \Bigg{(}    e^{(R_{1} + 1)X/2} -  e^{(R_{1} - 1)X/2 }  \Bigg{)}^{+}
	 e^{ -\frac{(x-X)^2}{4\tau} -\frac{(y-Y)^2}{4\tau}} dX dY,
\end{multline}
which leads to 
\begin{align}
\psi_{0}(x,y,\tau) &= \frac{1}{4 \pi\tau} \int \psi_{0}(X,Y,0) e^{  -\frac{(x-X)^2}{4\tau} -\frac{(y-Y)^2}{4\tau}} dX dY, 
\end{align}
and
\begin{multline}
{\psi}_{0}(x,y,\tau) = \frac{1}{4\pi\tau} \int e^{ (R_{2} - 1)Y/2} \Bigg{(}    e^{(R_{1} + 1)X/2 } -  e^{(R_{1} - 1)X/2 }  \Bigg{)}^{+}
	e^{ -\frac{(x-X)^2}{4\tau} -\frac{(y-Y)^2}{4\tau}} dX dY.
\end{multline}
Note that the two integrals can be separated:
\begin{multline}
{\psi}_{0}(x,y,\tau) = \frac{1}{\sqrt{4\pi\tau}} \int_0^{\infty} dX
 \Bigg{(}    e^{(R_{1} + 1)X/2 } -  e^{(R_{1} - 1)X/2 }  \Bigg{)}^{+}
	e^{ -\frac{(x-X)^2}{4\tau}} 
	\\ \times \frac{1}{\sqrt{4\pi\tau}} \int_{-\infty}^{\infty} dYe^{ (R_{2} - 1)Y/2} e^{ -\frac{(y-Y)^2}{4\tau}},
\end{multline}
where the first of these integrals precisely corresponds to the 1+1 dimensional Black Scholes model.
We thus finally obtain
\begin{multline} \label{PSI0}
	{\psi}_{0}(x,y,\tau) = e^{\frac{1}{2}(R_2-1)\left(\frac{\tau}{2}(R_2-1)+y\right)} 
	\Bigg{[}  e^{(R_{1} + 1)x/2 + (R_{1} + 1)^{2}\tau/4}\mathcal{N}(d_1) 
	- e^{(R_{1} - 1)x/2 + (R_{1} - 1)^{2}\tau/4}\mathcal{N}(d_2) 
	 \Bigg{]},
\end{multline}
where
\begin{align}
\mathcal{N}(d)=\frac{1}{\sqrt{2 \pi}} \int_{-\infty}^d \exp\left(\frac{-z^2}{2}\right) dz,
\end{align}
and
\begin{align}
d_{1} &= \frac{x}{\sqrt{2\tau}} + \frac{\sqrt{2\tau}}{2}(R_{1} + 1)=\frac{\log(S/K)+(r+\sigma^2/2)(T-t)}{\sigma \sqrt{T-t}}, \\
d_{2} &= \frac{x}{\sqrt{2\tau}} + \frac{\sqrt{2\tau}}{2}(R_{1} - 1) = d_1-\sigma\sqrt{T-t}.
\end{align}
We have obtained an analytical solution to the symmetrical model. Remarkably, because of the boundary condition that only depends on $S$, it is identical to the Black Scholes solution. Going back to the original variables we find:
\begin{equation}
	C_{0}(S,V,t) = S \mathcal{N}(d_1) -K e^{-r(T-t)}\mathcal{N}(d_2).
\end{equation}

\section{Symmetry Breaking terms and solution to the Merton Garman model}
\label{Symmetry Breaking terms and solution to the Merton Garman model}

In this Appendix, we give details of the derivation of the perturbative solution to the full Merton Garman. We first need to restore the full model by re-introducing the symmetry breaking terms
\begin{multline}
\label{fullM2}	\frac{\partial C}{\partial t} + rS\frac{\partial C}{\partial S} + \frac{1}{2}\sigma^{2}S^{2}\frac{\partial^{2}C}{\partial S^{2}} + \frac{c_{1}S^{2}}{2}\Bigg{(} V - \sigma^{2} \Bigg{)} \frac{\partial^{2}C}{\partial S^{2}} + \mu V\frac{\partial C}{\partial V} + c_{2}\lambda\frac{\partial C}{\partial V} + \frac{1}{2} \xi_{0}^{2}V^{2}\frac{\partial^{2}C}{\partial V^{2}}  
	\\ + \frac{1}{2} c_{3} \Bigg{(} \xi^{2}V^{2\alpha} - \xi_{0}^{2}V^{2}  \Bigg{)}\frac{\partial^{2} C}{\partial V^{2}} + c_{4}\rho\xi V^{\alpha + 1/2}S\frac{\partial^{2} C}{\partial S\partial V} - r C = 0.
\end{multline}
Note that we have introduced dimensionless coefficients $c_{i}$ which denote the strength of the symmetry breaking terms. In the limit $c_{i}=1$ one recovers the original Merton Garman model. These coefficients are simply introduced as a bookkeeping trick to keep track of which terms correspond to a deviation of the 2+1 Galilean invariant theory. In the end of the day, we set $c_{i}=1$. We can now apply the same variables transformations to Eq. (\ref{fullM2}) that we had applied in the symmetric case and obtain 
\begin{equation}
\left( \frac{\partial }{\partial \tau} -  \frac{\partial^2 }{\partial x^2}- \frac{\partial^2 }{\partial y^2}+\mathcal{D}(x,y)\right)\psi(x,y,\tau)=0,
\end{equation}
where the operator $\mathcal{D}(x,y)$ is defined as:
\begin{eqnarray}
\mathcal{D}(x,y) &=&
	\left ( \frac{c_1}{2}  \left ( V_0 e^y- \sigma^2 \right) \left ( (a^2-a)+(2a-1)\frac{\partial}{\partial x}+\frac{\partial^2}{\partial x^2} \right) \right . \\ \nonumber && \left . +  c_2 \frac{\lambda}{V_0} e^{-y} \left ( \frac{\partial}{\partial y} + b \right)
	\right . \\ \nonumber && \left . +c_3 \frac{1}{2} \left (\xi^2 V_0^{2\alpha-2} e^{(2\alpha-2)y}- \xi_0^2 \right)\left ( (b^2-b)+(2b-1)\frac{\partial}{\partial y}+\frac{\partial^2}{\partial y^2}\right)\right . \\ \nonumber && \left . 
	+c_4 \xi \rho V_0^{\alpha-\frac{1}{2}} e^{\left(\alpha-\frac{1}{2}\right)y} 
	\left(a b+ b \frac{\partial}{\partial x}+a\frac{\partial}{\partial y} +\frac{\partial^2}{\partial x\partial y} \right)
	 \right),
\end{eqnarray}
where $a$ and $b$ are respectively given in Eq. (\ref{a}) and Eq. (\ref{b}). Note that $\mathcal{D}(x,y)$ is $\tau$ independent.

We now do perturbation theory around the symmetrical solution $\psi_{0}$ which was given in Eq. (\ref{PSI0}). To leading order in $c_i$, we write $\psi=\psi_{0}+\psi_{1} $ where $\psi_{1}$ is of order $c_i$. Keeping in mind that $\mathcal{D}$ is order $c_i$, we find
\begin{equation}
	\left( \frac{\partial }{\partial \tau} -  \frac{\partial^2 }{\partial x^2}- \frac{\partial^2 }{\partial y^2}\right) \psi_{1}(x,y,\tau) = -\mathcal{D}(x,y)\psi_{0}(x,y,\tau).
\end{equation}
This equation can be solved by the Green's function method, we obtain
\begin{equation}
\psi_1(x,y,\tau)= - \int_0^\tau dt \int_{-\infty}^\infty dx^\prime \int_{-\infty}^\infty  dy^\prime G(x,y,\tau;x^\prime,y^\prime, t) \mathcal{D}(x^\prime,y^\prime) \psi_{0}(x^\prime,y^\prime, t),
\end{equation}
where
\begin{equation}
G(x,y,\tau;x^\prime,y^\prime, t) = \frac{1}{4 \pi (\tau-t)}
 \exp\left(- \frac{(x-x^\prime)^2+(y-y^\prime)^2}{4(\tau-t)}\right).
\end{equation}
These integrals can be performed analytically. Our result provides an approximative and analytical solution to the Merton Garman model. We find:
\begin{equation}
\psi(x,y,\tau)=\psi_0(x,y,\tau)+\psi_1(x,y,\tau),
\end{equation}
with
\begin{equation} \label{cor1}
\psi_1(x,y,\tau)=
\frac{c_{1} \left(\sigma ^2 \tau \left(\sqrt{2}\gamma +\sigma \frac{\xi_{0}}{\sqrt{2}}  \right)-\sigma \frac{\xi_{0}}{\sqrt{2}}  V_{0} e^y \left(e^{\frac{2\gamma  \tau}{\sigma \xi_{0}}+\tau}-1\right)\right)e^{\left(\frac{\gamma ^2 \tau}{\sigma \xi_{0}^2}-\frac{x^2}{4 \tau}+\frac{\gamma  y}{\sigma \xi_{0}}\right)}}{4 \sqrt{\pi\tau }  \left(\sqrt{2} \gamma +\sigma \frac{\xi_{0}}{\sqrt{2}}  \right)},
   \end{equation}
to leading order. In the original variables, we find:
\begin{eqnarray} 
C_1(S,V,t)&=&-K  \frac{\left(\frac{S}{K}\right)^{\frac{1}{2}-\frac{r}{\sigma ^2}} e^{\left(\frac{4 \log
   ^2\left(\frac{S}{K}\right)+\left(2 r+\sigma ^2\right)^2 (t-T)^2}{8 \sigma ^2
   (t-T)}\right)} }{4 \sqrt{2 \pi } \left(\frac{2 \gamma }{\sigma \xi_{0}}+1\right) \sqrt{\sigma ^2 (T-t)}} \\ \nonumber 
   &&  \times \left(\frac{1}{2} \sigma ^4 \left(\frac{2 \gamma }{\sigma \xi_{0}}+1\right) (t-T)+V \left(e^{\frac{1}{2} \sigma ^2 \left(\frac{2 \gamma }{\sigma \xi_{0}}+1\right) (T-t)} - 1 \right)\right),
 \end{eqnarray}
where we have set $c_1=1$.

It may appear surprising that the leading order correction does not depend on the symmetry breaking terms parametrized by $c_2$, $c_3$ and $c_4$.  To understand what is happening, one can organize the perturbation theory slightly different, but mathematically equivalent, fashion by looking at corrections to the Green's function $G(x,y,\tau | x^\prime,y^\prime,0)$. The differential equation to solve is given by:
\begin{equation}
\left( \frac{\partial }{\partial \tau} -  \frac{\partial^2 }{\partial x^2}- \frac{\partial^2 }{\partial y^2}+\mathcal{D}\right)G(x,y,\tau | x^\prime,y^\prime,0)
	=\delta(\tau) \delta(x-x^\prime) \delta(y-y^\prime).
\end{equation}
 Perturbation theory is organized by expanding around the Green's function of the symmetrical symmetry:
 $G(x,y,\tau | x^\prime,y^\prime,0)=G_0(x,y,\tau | x^\prime,y^\prime,0)+G_1(x,y,\tau | x^\prime,y^\prime,0)$ to leading order.
 One obtains:
 \begin{equation}
 \left( \frac{\partial }{\partial \tau} -  \frac{\partial^2 }{\partial x^2}- \frac{\partial^2 }{\partial y^2}\right)G_0(x,y,\tau | x^\prime,y^\prime,0)
	=\delta(\tau) \delta(x-x^\prime) \delta(y-y^\prime).
 \end{equation}
The solution to this partial differential equation was given above. The correction $G_1(x,y,\tau | x^\prime,y^\prime,0)$ is obtained by solving:
 \begin{equation}
 \left( \frac{\partial }{\partial \tau} -  \frac{\partial^2 }{\partial x^2}- \frac{\partial^2 }{\partial y^2}\right)G_1(x,y,\tau | x^\prime,y^\prime,0)
	=-\mathcal{D}(x,y) G_0(x,y,\tau | x^\prime,y^\prime,0),
 \end{equation}
which can be solved easily. One finds:
 \begin{equation}
G_1(x,y,\tau | x^\prime,y^\prime,0)
	=-\int_0^\tau dt^\prime \int_{-\infty}^\infty dx^{\prime\prime} \int_{-\infty}^\infty dy^{\prime\prime}
	G_0(x,y,\tau | x^{\prime\prime},y^{\prime\prime},t^\prime)\mathcal{D}(x^{\prime\prime},y^{\prime\prime}) G_0(x^{\prime\prime},y^{\prime\prime},t^\prime | x^{\prime},y^{\prime},0).
 \end{equation}
 It is easy to show that this correction depends on all four symmetry breaking terms. We find
\begin{eqnarray}
G_{1,c_1}(x,y,\tau | x^\prime,y^\prime,0)&=&
 \frac{c_1}{64 \pi \tau^{5/2}}e^{-\frac{(x-x^\prime)^2+(y-y^\prime)^2}{4 \tau}}
 \left ((-1+R_1)^2  \tau^2+2 \tau (-1+R_1(x-x^\prime))+(x-x^\prime)^2\right )  \nonumber \\ && \nonumber
  \times \left( 2 \sigma^2\sqrt{\tau}-\sqrt{\pi}e^{\frac{\tau^2+(y-y^\prime)^2+2 \tau (y-y^\prime)}{4 \tau}} V_0
   \right . 
    \nonumber \\ && \nonumber \left .  
   \left(Erf\left(\frac{\tau+y-y^\prime}{2\sqrt{\tau}}\right)
  +Erf\left(\frac{\tau-y+y^\prime}{2\sqrt{\tau}}\right)\right)\right), \\ &&
\end{eqnarray}
\begin{eqnarray}
G_{1,c_2}(x,y,\tau | x^\prime,y^\prime,0)&=&
c_2 \frac{\lambda}{16 \pi V_0 \tau^{2}}e^{-\frac{(x-x^\prime)^2+y^2+y^{\prime 2}+4 \tau (y-y^\prime)}{4 \tau}}  \nonumber \\ && \nonumber
\Bigg ( 2 e^{y y^\prime}{2 \tau}(e^y-e^{y^\prime}) \tau + e^{(\tau+y)^2+2 \tau y^\prime+y^{\prime 2}}{4 \tau} \sqrt{\pi} (R_2-2) \tau^{3/2}  \nonumber \\ &&
\left( Erf\left(\frac{\tau-y+y^\prime}{2\sqrt{\tau}}\right)-Erf\left(\frac{-\tau-y+y^\prime}{2\sqrt{\tau}}\right)\right)\Bigg),       \end{eqnarray}
 \begin{eqnarray}
 G_{1,c_3}(x,y,\tau | x^\prime,y^\prime,0)&=&
 c_3\frac{1}{64 \tau^2\pi(1-\alpha)} e^{-\frac{(x-x^\prime)^2}{4 \tau}}  \\ &&
 \times \Bigg ( 2 e^{-\frac{(y-y^\prime)^2}{4 \tau}-2 (x-x^\prime)}
 (-2 (\alpha -1) \xi_0^2  e^{2 (y+y^\prime)} 
 \nonumber \\ && \nonumber
   \left(\left(R_2^2-1\right) \tau^2+2 \tau (R_2 y-R_2
   y^\prime-1)+(y-y^\prime)^2\right)
   \nonumber \\ && \nonumber
   +\xi^2 V^{2 \alpha-2 }
   \left(-e^{2 (\alpha  y+y^\prime)}\right) (-2 \tau (\alpha
   +R_2-1)-y+y^\prime)\nonumber \\ && \nonumber
   -\xi ^2 V^{2 \alpha-2 } e^{2 (y+\alpha 
   y^\prime)} (2 \tau (3 \alpha +R_2-3)+y-y^\prime))
   \nonumber \\ && \nonumber
   +\sqrt{\pi } \xi ^2 \tau^{3/2} V^{2 \alpha-2 } (2 \alpha +R_2-3) (2
   \alpha +R_2-1) e^{(\alpha -1) ((\alpha -1) \tau+y+y^\prime)}
   \nonumber \\ && \nonumber
   \left(Erf\left(\frac{2 (\alpha -1) \tau+y-y^\prime}{2
   \sqrt{\tau}}\right)-Erf\left(\frac{-2 (\alpha -1) \tau+y-y^\prime}{2
   \sqrt{\tau}}\right)\right)
 \Bigg ),
      \end{eqnarray}
      and
 \begin{eqnarray}
 G_{1,c_4}(x,y,\tau | x^\prime,y^\prime,0)&=&
 c_4
  \frac{\xi  \rho  V_0^{\alpha -\frac{1}{2}} ((R_1-1) \tau+x-x^{\prime})}{32 \pi \tau^2 (1- 2 \alpha)}
   e^{-\frac{\alpha  \tau^2+y (\tau+y)+(x-x^{\prime})^2+y^{\prime 2}}{4 \tau}}
\\ && \nonumber
    \Bigg(4
   \left(e^{\alpha  y+\frac{y^{\prime}}{2}}-e^{\frac{y}{2}+\alpha  y^{\prime}}\right)
   e^{\frac{1}{4} \left(\alpha  \tau+y \left(\frac{2 y^{\prime}}{\tau}-1\right)
   -2 y^{\prime}\right)}
   \nonumber \\ && \nonumber
   -\sqrt{\pi  \tau} (2 \alpha +2 R_2-3)
   \nonumber \\ && \nonumber
   \left(Erf\left(\frac{-2 \alpha  \tau+\tau+2 y-2 y^{\prime}}{4
   \sqrt{\tau}}\right)-Erf\left(\frac{2 \alpha  \tau-\tau+2 y-2 y^{\prime}}{4
   \sqrt{\tau}}\right)\right)
   \nonumber \\ && \nonumber
    \exp \left(\frac{4 \alpha ^2 \tau^2+8 \alpha  \tau
   (y+y^{\prime})+(\tau-2 y^{\prime})^2+4 y^2}{16 \tau}\right)\Bigg).
 \end{eqnarray}
 
 It is easy to see that when folding these corrections to the Green's function with the boundary condition  (\ref{transformed BC}) that only the contribution from the $c_1$ term survives and one recovers the result of (\ref{cor1}). The boundary condition thus implies that the contributions of $c_2$, $c_3$ and $c_4$ vanish to leading order in the perturbation theory. These symmetry breaking terms will, however, contribute to higher order corrections. Higher precision, if required, can be obtained by going to higher order in perturbation theory. Option prices can be calculated extremely rapidly using this formalism. Note that, in principle, if we resummed perturbation theory to all order in $c_i$, the dependence on $\sigma$ and $\xi_0$ would  vanish. It is also worth noticing that our results are independent on $V_0$ which is only introduced to match the dimension of $V$. 

\section{Time Series Simulation}
\label{appendix: time series}
The Merton Garman model is defined by the coupled two-dimensional SDE:
\begin{align}
	dS_{t} &= rS_{t}dt + \sqrt{V_{t}}S_{t}dW_{t}^{S}, \\
	dV_{t} &= \kappa(\theta - V_{t})dt + \xi V_{t}^{\alpha}dW_{t}^{V},
\end{align}
with the two Brownian motion components: $W_{t}^{S},W_{t}^{V}$ having correlation $\rho$. We use an Euler discretization scheme for the asset price and variance process, with a full truncation to tackle the issue of negative variances. Conditional on a time $s$ for $t>s$ the discretization scheme for the asset price and variance processes are:
\begin{align}
S_{t} &= S_{s} + rS_{t}\Delta t + S_{s}\sqrt{V_{t}^{+}\Delta t}z_{S}, \\
V_{t} &= V_{s} - \kappa(\theta - V_{s}^{+})\Delta t + \xi (V_{s}^{+}\Delta t)^{\alpha}z_{V},
\end{align}
where $\Delta t = t-s$, $z_{V} \sim \mathcal{N}(0,1)$ and $z_{S} = \rho z_{V} + \sqrt{1-\rho}z$ with $z\sim \mathcal{N}(0,1)$. This scheme is used to generate 100 different sample paths of weekly returns and latent variance over one year, i.e 52 observations with $\Delta t = 7$ days. At each observation time we simulate six unique maturities within $[7,180]$ days maturity, with a moneyness range of $K/S\in [0.9,1.1]$ across ten strikes. Each option price is computed using the Monte Carlo framework with $50,000$ simulations and a time-step of ${1/20}^{\text{th}}$ of a trading day. 

The calibration process is done using the objective function defined in Eq. (\ref{IV objective function}). It should be noted that as initial conditions we start with the true parameter vector, i.e $\Theta^{\text{pert.}}_{\text{initial}}=\Theta^{MG}_{\text{true}}$ and for $\sigma$ we start with the initial variance. We also pass the variance path at each time step.

\clearpage

\newpage
\begin{longtable}{ccccccccccccccccccccccccccccccccccc}
\caption{Implied Volatility Root Mean Squared Error (IVRMSE, defined in Eq. (\ref{IV objective function})) when calibrated the leading order order perturbative solution and compared to the Monte Carlo simulation of the Merton Garman model for four different initial volatilities for a time horizon of 30 days. Also reported is the calibrated values of $\sigma$ for each scenario.}

\\ \hline
&$V$ &&&&&&&&&&&&& $\sigma$ &&&&&&&&&&&&&& $\text{IVRMSE}$ &
 \\ \hline 

&0.3500&&&&&&&&&&&&&0.3254&&&&&&&&&&&&&&0.0055& \\ 
&0.2500&&&&&&&&&&&&&0.2394&&&&&&&&&&&&&&0.0037& \\
&0.1800&&&&&&&&&&&&&0.1743&&&&&&&&&&&&&&0.0070& \\ 
&0.1000&&&&&&&&&&&&&0.1069&&&&&&&&&&&&&&0.0157& \\ 
\hline 
\label{tab: static}
\end{longtable}

\newpage
\begin{table}
\caption{Simulated calibration exercise for the leading order order perturbative solution with 100 paths, per data set. The MG row displays the true parameter vector. The leading order perturbative solution (Pert.) row displays the mean calibrated results of the leading order order perturbative solution. The std. error row displays the stand deviation of the leading order order perturbative solution parameters.}
\begin{tabular}{c c c c c c c c c c c c c}
\\ 
  &&& \text{Panel 1. Dataset 1}  \\ 
&& $\kappa$& $\theta$ & $\xi$& &$\rho$&&$\alpha$&&$\sigma^2$ \\ \hline 
MG&&1.1768 & 0.0823 & 0.3000 && -0.5459 && 1.0000  &&- \\ 
Pert. && 4.1195 & - & 0.7865 && - && 0.9998 && 0.0809 \\ 
Bias&&2.9427 & - & 0.4865  && - && 0.0002 && - \\ 
Std. Error && 3.3627 & - & 0.4054 && - && 0.1605 && 0.0125\\
\hline
\\ 
  &&& \text{Panel 2. Dataset 2}  \\ 
&& $\kappa$& $\theta$ & $\xi$& &$\rho$&&$\alpha$&&$\sigma^2$ \\ \hline 
MG &&1.1768 & 0.0823 & 0.3000 && 0.0000 && 1.0000 && -  \\ 
Pert. && 2.3869 & - & 0.6012 &&- && 0.6219 && 0.0808 \\ 
Bias&&1.2101 & - & 0.3012 &&- && 0.3781 && - \\ 
Std. Error && 3.2793 & - & 0.4812  && - && 0.3219 && 0.0137 \\ 
\hline
\\ 
  &&& \text{Panel 3. Dataset 3}  \\ 
&& $\kappa$& $\theta$ & $\xi$& &$\rho$&&$\alpha$&&$\sigma^2$ \\ \hline  
MG & & 1.1768 & 0.0823 & 0.3000  & &0.5459 & & 1.0000 & & - \\
Pert. & & 4.3731 &- & 0.7939 & &- & & 0.9785 & & 0.0801 \\
Bias& &3.1963 & - & 0.4939 & & - & & 0.0215 & & - \\ 
Std. Error & & 3.1934 & - & 0.3034  &&- && 0.1269 && 0.0124\\ 
\hline
\\ 
  &&& \text{Panel 4. Dataset 4}  \\  
  && $\kappa$& $\theta$ & $\xi$& &$\rho$&&$\alpha$&&$\sigma^2$ \\ \hline 
  MG & & 1.1768 & 0.1250 & 0.3000 & & -0.5459 & & 1.0000 & & - \\
Pert. & & 4.2349 & - & 0.9737 &&  - & & 1.6660 & & 0.1016 \\
Bias& &3.0581 & - & 0.6737  & & - & & 0.6660 & & - \\  
Std. Error& &4.0018 & - & 0.4312 & & - & & 1.4387 & & 0.0142 \\ 
\hline 
\end{tabular}
\label{tab: parameter results} 
\end{table}

\newpage
\begin{table}
\caption{Displays IVRMSE between the perturbative solution (Pert.) and the solution of the Merton Garman model for each data set. IVRMSE is calculated using 23. The std. error row denotes the standard deviation of the IVRMSE.}
\begin{tabular}{c c c c c c c c c c c c c c c c c c} 
    \\ 
& &$\text{data set 1}$& &  &  & $\text{data set 2}$ & &  &  & $\text{data set 3}$& &  &  & $\text{data set 4}$& \\ \hline 
\text{IVRMSE}&&0.0128 &  &  &  &  0.0116 &  &  &  &  0.0131 &  &  &  &  0.0170\\
\text{Std. Error} && 0.0034 & &  &  &  0.0030 &  &  &  &   0.0035 & &  &  & 0.0059\\
\hline 
\end{tabular} 
\label{tab: error results}
\end{table}

%%%%%%%%%%%%%%%%%%%%%%%%%%%%%%%%%%%%%%%%%%%%%%%%%%%%%%%%%%%%%%%%%
%%%
%%%                     BIBLIOGRAPHy
%%%
%%%%%%%%%%%%%%%%%%%%%%%%%%%%%%%%%%%%%%%%%%%%%%%%%%%%%%%%%%%%%%%%%

\clearpage 
\nocite{*}
\bibliographystyle{acm}
\bibliography{biblio}

\begin{thebibliography}{10}

\bibitem{aguilar2017series}
{\sc Aguilar, J.-P.}
\newblock A series representation for the black-scholes formula.
\newblock {\em arXiv preprint arXiv:1710.01141\/} (2017).

\bibitem{ai2007maximum}
{\sc A{\i}, Y., Kimmel, R., et~al.}
\newblock Maximum likelihood estimation of stochastic volatility models.
\newblock {\em Journal of financial economics 83}, 2 (2007), 413--452.

\bibitem{ajaib2017introducing}
{\sc Ajaib, M.~A.}
\newblock Introducing spin in 2d quantum tunneling.
\newblock {\em arXiv preprint arXiv:1705.10409\/} (2017).

\bibitem{andersen2002empirical}
{\sc Andersen, T.~G., Benzoni, L., and Lund, J.}
\newblock An empirical investigation of continuous-time equity return models.
\newblock {\em The Journal of Finance 57}, 3 (2002), 1239--1284.

\bibitem{baaquie1997path}
{\sc Baaquie, B.~E.}
\newblock A path integral approach to option pricing with stochastic
  volatility: some exact results.
\newblock {\em Journal de Physique I 7}, 12 (1997), 1733--1753.

\bibitem{baaquie2003quantum}
{\sc Baaquie, B.~E., Coriano, C., and Srikant, M.}
\newblock Quantum mechanics, path integrals and option pricing: Reducing the
  complexity of finance.
\newblock In {\em Nonlinear Physics: Theory and Experiment II}. World
  Scientific, 2003, pp.~333--339.

\bibitem{bachelier1900theorie}
{\sc Bachelier, L.}
\newblock {\em Th{\'e}orie de la sp{\'e}culation}.
\newblock Gauthier-Villars, 1900.

\bibitem{bakshi1997empirical}
{\sc Bakshi, G., Cao, C., and Chen, Z.}
\newblock Empirical performance of alternative option pricing models.
\newblock {\em The Journal of finance 52}, 5 (1997), 2003--2049.

\bibitem{bardgett2018inferring}
{\sc Bardgett, C., Gourier, E., and Leippold, M.}
\newblock Inferring volatility dynamics and risk premia from the s\&p 500 and
  vix markets.
\newblock {\em Journal of Financial Economics\/} (2018).

\bibitem{bates2000post}
{\sc Bates, D.~S.}
\newblock Post-'87 crash fears in the s\&p 500 futures option market.
\newblock {\em Journal of Econometrics 94}, 1-2 (2000), 181--238.

\bibitem{bates2006maximum}
{\sc Bates, D.~S.}
\newblock Maximum likelihood estimation of latent affine processes.
\newblock {\em The Review of Financial Studies 19}, 3 (2006), 909--965.

\bibitem{benzoni2002pricing}
{\sc Benzoni, L.}
\newblock Pricing options under stochastic volatility: an empirical
  investigation.
\newblock Tech. rep., working paper, University of Minnesota, 2002.

\bibitem{black1973pricing}
{\sc Black, F., and Scholes, M.}
\newblock The pricing of options and corporate liabilities.
\newblock {\em Journal of political economy 81}, 3 (1973), 637--654.

\bibitem{blazhyevskyi2011path}
{\sc Blazhyevskyi, L., and Yanishevsky, V.}
\newblock The path integral representation kernel of evolution operator in
  merton-garman model.
\newblock {\em arXiv preprint arXiv:1106.5143\/} (2011).

\bibitem{bose1995representations}
{\sc Bose, S.}
\newblock Representations of the (2+ 1)-dimensional galilean group.
\newblock {\em Journal of Mathematical Physics 36}, 2 (1995), 875--890.

\bibitem{chernov2003alternative}
{\sc Chernov, M., Gallant, A.~R., Ghysels, E., and Tauchen, G.}
\newblock Alternative models for stock price dynamics.
\newblock {\em Journal of Econometrics 116}, 1-2 (2003), 225--257.

\bibitem{chourdakis2011maximum}
{\sc Chourdakis, K., and Dotsis, G.}
\newblock Maximum likelihood estimation of non-affine volatility processes.
\newblock {\em Journal of Empirical Finance 18}, 3 (2011), 533--545.

\bibitem{christoffersen2014economic}
{\sc Christoffersen, P., Feunou, B., Jacobs, K., and Meddahi, N.}
\newblock The economic value of realized volatility: Using high-frequency
  returns for option valuation.
\newblock {\em Journal of Financial and Quantitative Analysis 49}, 3 (2014),
  663--697.

\bibitem{christoffersen2010volatility}
{\sc Christoffersen, P., Jacobs, K., and Mimouni, K.}
\newblock Volatility dynamics for the s\&p500: evidence from realized
  volatility, daily returns, and option prices.
\newblock {\em The Review of Financial Studies 23}, 8 (2010), 3141--3189.

\bibitem{duan2010jump}
{\sc Duan, J.-C., and Yeh, C.-Y.}
\newblock Jump and volatility risk premiums implied by vix.
\newblock {\em Journal of Economic Dynamics and Control 34}, 11 (2010),
  2232--2244.

\bibitem{eraker2004stock}
{\sc Eraker, B.}
\newblock Do stock prices and volatility jump? reconciling evidence from spot
  and option prices.
\newblock {\em The Journal of Finance 59}, 3 (2004), 1367--1403.

\bibitem{eraker2003impact}
{\sc Eraker, B., Johannes, M., and Polson, N.}
\newblock The impact of jumps in volatility and returns.
\newblock {\em The Journal of Finance 58}, 3 (2003), 1269--1300.

\bibitem{garman1976general}
{\sc Garman, M.~B., et~al.}
\newblock A general theory of asset valuation under diffusion state processes.
\newblock Tech. rep., University of California at Berkeley, 1976.

\bibitem{heston1993closed}
{\sc Heston, S.~L.}
\newblock A closed-form solution for options with stochastic volatility with
  applications to bond and currency options.
\newblock {\em The review of financial studies 6}, 2 (1993), 327--343.

\bibitem{huang2004specification}
{\sc Huang, J.-z., and Wu, L.}
\newblock Specification analysis of option pricing models based on time-changed
  l{\'e}vy processes.
\newblock {\em The Journal of Finance 59}, 3 (2004), 1405--1439.

\bibitem{hull1987pricing}
{\sc Hull, J., and White, A.}
\newblock The pricing of options on assets with stochastic volatilities.
\newblock {\em The journal of finance 42}, 2 (1987), 281--300.

\bibitem{hull1997options}
{\sc Hull, J.~C.}
\newblock Options, futures and other derivatives. -prentice hall, upper saddle
  river, new jersey.

\bibitem{jones2003dynamics}
{\sc Jones, C.~S.}
\newblock The dynamics of stochastic volatility: evidence from underlying and
  options markets.
\newblock {\em Journal of econometrics 116}, 1-2 (2003), 181--224.

\bibitem{kaeck2012volatility}
{\sc Kaeck, A., and Alexander, C.}
\newblock Volatility dynamics for the s\&p 500: Further evidence from
  non-affine, multi-factor jump diffusions.
\newblock {\em Journal of Banking \& Finance 36}, 11 (2012), 3110--3121.

\bibitem{kleinert2016option}
{\sc Kleinert, H., and Korbel, J.}
\newblock Option pricing beyond black--scholes based on double-fractional
  diffusion.
\newblock {\em Physica A: Statistical Mechanics and its Applications 449\/}
  (2016), 200--214.

\bibitem{levy1967nonrelativistic}
{\sc L{\'e}vy-Leblond, J.-M.}
\newblock Nonrelativistic particles and wave equations.
\newblock {\em Communications in Mathematical Physics 6}, 4 (1967), 286--311.

\bibitem{merton1973theory}
{\sc Merton, R.~C.}
\newblock Theory of rational option pricing.
\newblock {\em The Bell Journal of economics and management science\/} (1973),
  141--183.

\bibitem{pan2002jump}
{\sc Pan, J.}
\newblock The jump-risk premia implicit in options: Evidence from an integrated
  time-series study.
\newblock {\em Journal of financial economics 63}, 1 (2002), 3--50.

\bibitem{park2015price}
{\sc Park, Y.-H.}
\newblock Price dislocation and price discovery in the s\&p 500 options and vix
  derivatives markets.
\newblock Tech. rep., Working Paper). Federal Reserve Board, 2015.

\bibitem{utama2016feynman}
{\sc Utama, B., and Purqon, A.}
\newblock Feynman path integral application on deriving black-scholes diffusion
  equation for european option pricing.
\newblock In {\em Journal of Physics: Conference Series\/} (2016), vol.~739,
  IOP Publishing, p.~012021.

\end{thebibliography}

\end{document}